\documentclass[fleqn,11pt]{article}

\usepackage[affil-it]{authblk}
\usepackage[english]{babel}
\usepackage{amsmath,amssymb,amsfonts,amsthm}
\usepackage[hidelinks]{hyperref}
\hypersetup{colorlinks, linkcolor=blue, citecolor=blue, urlcolor=blue}
\usepackage{cite}
\usepackage{mathrsfs}

\usepackage[title,titletoc,toc]{appendix}
\usepackage[margin=1in,left=1in,right=1in]{geometry}
\usepackage{longtable,array}
\usepackage{booktabs}
\usepackage{subcaption}
\usepackage{tabularx}

\theoremstyle{definition}

\usepackage{rotating,graphicx}

\providecommand{\keywords}[1]{\textbf{\textit{Keywords:}} #1}
\allowdisplaybreaks

\title{On the equivalence of three extended Calogero--Bogoyavlenskii--Schiff models to the classical CBS equation}

\author[1]{Manjit Singh\thanks{Email: manjitcsir@pbi.ac.in, manjitcsir@gmail.com}}
\affil[1]{Yadavindra Department of Sciences, Punjabi University Guru Kashi Campus, Talwandi Sabo--151302, Punjab, India.}
\date{}
\begin{document}
\maketitle
\begin{abstract}
We examine the three extended Calogero--Bogoyavlenskii--Schiff
(CBS) models introduced by Wazwaz, Alhejaili, and El-Tantawy
[Eur.\ Phys.\ J.\ Plus \textbf{141}, 1218 (2026)].
Explicit invertible point transformations map the first
model, and generically the second, to the classical CBS
equation. The third model is generically equivalent to CBS
with an additional spectator coordinate absent from the
differential operator. We identify the exceptional parameter
regimes and explain the reported exponential dispersion
relations through these transformations. These equivalences show that the proposed models do not,
in the generic parameter regimes, introduce new
point-inequivalent CBS equations or genuine
four-variable differential coupling.
\end{abstract}

\section{Introduction}\label{sec:introduction}

In a recent article, Wazwaz, Alhejaili, and El-Tantawy
\cite{Wazwaz2026ThreeCBS} introduced three extended
Calogero--Bogoyavlenskii--Schiff (CBS) models: two in
$(2+1)$ dimensions and one in $(3+1)$ dimensions.
The proposed equations combine the classical CBS nonlinear
structure with additional linear derivative terms and,
in the third model, a second CBS-type block involving
the original time variable. The authors investigated
these equations using the Weiss--Tabor--Carnevale
Painlev\'e test, exponential multi-wave constructions,
and traveling-wave reductions. They presented the models
as new integrable extensions, including a higher-dimensional
equation with nontrivial transverse and temporal coupling.

Before interpreting additional terms or independent
variables as evidence of a new integrable structure,
one must determine whether they can be eliminated by
invertible point transformations. Equations with different
appearances may represent the same differential equation
in different coordinates. Such equivalences transport
local solutions and auxiliary integrability structures
and therefore provide a necessary basis for assessing
the novelty of a proposed extension.

The reference equation in this comment is the classical
potential CBS equation, written in the normalization
\begin{equation}\label{eq:canonical-CBS}
 v_{XT}+v_{XXXY}
 +\frac12v_Yv_{XX}+v_Xv_{XY}=0.
\end{equation}
This normalization follows from the potential CBS equation
in \cite[Eq.~(17), p.~594]{Wazwaz2008Multiple}
by multiplying its dependent variable by four.
It also occurs directly in the general equation of
\cite[Eq.~(19), p.~174]{Wazwaz2010CBS}
when the nonlinear coefficients are $a=1$ and
$b=\tfrac12$. Earlier exact-solution treatments include
\cite{LiChen2004}, while the logarithmic multi-wave
constructions relevant here appear in
\cite[Sec.~3, pp.~594--595]{Wazwaz2008Multiple}
and
\cite[Sec.~3.1, p.~175]{Wazwaz2010CBS}.

We establish explicit invertible point transformations
relating the three models of \cite{Wazwaz2026ThreeCBS}
to \eqref{eq:canonical-CBS}. The first model is
point-equivalent to this equation for all values of its
two parameters, and the second has the same property
when $\alpha\neq0$. For the third model, whenever
$(1-\alpha,\gamma)\neq(0,0)$, the transformed equation
is \eqref{eq:canonical-CBS} with an additional independent
variable $R$ absent from the differential operator.
Thus, it describes a family of CBS solutions parameterized
by $R$, without differential coupling in that direction.
This conclusion follows from an invertible transformation
of the full equation and does not impose $v_R=0$.
The exceptional parameter regimes are treated separately.

The solution comparison goes beyond equation equivalence.
Using the same transformations, together with an
affine-background adjustment where required, we recover
the reported one-, two-, and three-wave families from
the earlier constructions in
\cite[Eqs.~(24)--(31), pp.~594--595]{Wazwaz2008Multiple}
and
\cite[Eqs.~(27)--(35), p.~175]{Wazwaz2010CBS}.
The transformations account for the parameter-dependent
exponential dispersion relations while preserving the
pairwise interaction coefficients. Moreover, the
four-variable equation in
\cite[Eq.~(63), p.~177]{Wazwaz2010CBS},
for $a=c=4$ and $b=d=2$, also admits a CBS
representation with a spectator coordinate, as shown
below.

These results require a reassessment of the claimed
novelty of the models and their higher-dimensional
coupling. They explain both the underlying CBS structure
and the recovery of the specified multi-wave families
from earlier formulas. They do not establish that every
traveling-wave or cnoidal expression reported in
\cite{Wazwaz2026ThreeCBS} has previously appeared,
nor do they invalidate solutions merely because those
solutions are expressed in noncanonical coordinates.

Our conclusions concern local differential equations.
Transformations that interchange independent variables
or introduce affine backgrounds need not preserve a
prescribed physical initial--boundary value problem.
Any distinct physical interpretation therefore requires
a separate formulation of that problem. We first present
the equivalence transformations and their parameter
restrictions, then establish the solution correspondence
and discuss its implications for integrability and
dimensional novelty.
\section{Equivalence of the two $(2+1)$-dimensional models}
\label{sec:two-models}

We use explicit point transformations in the standard sense
\cite{olverbook}. Throughout, the transformed equation is
\eqref{eq:canonical-CBS}, unless an exceptional parameter
regime is specified.

\subsection{First model: removal of constant drift terms}
\label{subsec:first-model}

The first model, equation~(4) of
\cite{Wazwaz2026ThreeCBS}, is
\begin{equation}\label{eq:first-model}
 u_{xt}+u_{xxxy}+\frac12u_yu_{xx}+u_xu_{xy}
 +\alpha u_{xy}+\beta u_{xx}=0.
\end{equation}
The translating coordinate transformation
\begin{equation}\label{eq:first-transformation}
 \begin{gathered}
 X=x-\beta t,\qquad Y=y-\alpha t,\qquad T=t,\\
 u(x,y,t)=v(X,Y,T)
 \end{gathered}
\end{equation}
has unit Jacobian and maps \eqref{eq:first-model} exactly to
\eqref{eq:canonical-CBS}. Its inverse is
$x=X+\beta T$, $y=Y+\alpha T$, $t=T$.
Thus, for every $(\alpha,\beta)\in\mathbb{R}^2$, the added
terms are removable constant drifts and do not define a
point-inequivalent extension of CBS.

\subsection{Second model: affine change of variables}
\label{subsec:second-model}

The second model, equation~(5) of
\cite{Wazwaz2026ThreeCBS}, is
\begin{equation}\label{eq:second-model}
 u_{xt}+u_{xxxt}+\frac12u_tu_{xx}+u_xu_{xt}
 +\alpha u_{xy}+\beta u_{xx}=0.
\end{equation}
For $\alpha\neq0$, the point transformation
\begin{equation}\label{eq:second-transformation}
 \begin{gathered}
 X=x,\qquad Y=t,\qquad T=\frac{y}{\alpha},\\
 u(x,y,t)=v(X,Y,T)-x-2\beta t
 \end{gathered}
\end{equation}
maps \eqref{eq:second-model} exactly to
\eqref{eq:canonical-CBS}. The independent-variable Jacobian
is $-1/\alpha$, and the inverse is
\begin{equation}\label{eq:second-inverse}
 \begin{gathered}
 x=X,\qquad y=\alpha T,\qquad t=Y,\\
 v(X,Y,T)=u(X,\alpha T,Y)+X+2\beta Y.
 \end{gathered}
\end{equation}
The affine terms remove the separate $u_{xt}$ and
$\beta u_{xx}$ contributions; interchanging and rescaling
the independent variables supplies the canonical CBS
evolution term.

For $\alpha=0$, use instead $X=x$, $Y=t$, $R=y$ and
$u=v(X,Y,R)-x-2\beta t$. The resulting equation is
\begin{equation}\label{eq:second-degenerate}
 v_{XXXY}+\frac12v_Yv_{XX}+v_Xv_{XY}=0,
\end{equation}
with no $R$-derivatives. This exceptional regime is the
stationary CBS equation with one spectator coordinate.
Thus, the first model and the second model with
$\alpha\neq0$ are representations of the same classical
CBS equation.

\section{The $(3+1)$-dimensional model and its spectator coordinate}
\label{sec:third-model}

The third model, equation~(6) of
\cite{Wazwaz2026ThreeCBS}, is
\begin{equation}\label{eq:third-model}
\begin{aligned}
 0={}&u_{xt}+u_{xxxy}+u_{xxxt}
 +\frac12(u_y+u_t)u_{xx}\\
 &+u_x(u_{xy}+u_{xt})
 +\alpha u_{xy}+\beta u_{xx}+\gamma u_{xz}.
\end{aligned}
\end{equation}
Its two CBS-type blocks involve the same directional
derivative $\partial_y+\partial_t$. To expose this structure,
set
\begin{equation}\label{eq:third-intermediate-transformation}
 \begin{gathered}
 X=x,\qquad Y=y,\qquad S=t-y,\qquad Z=z,\\
 u(x,y,z,t)=w(X,Y,S,Z)-\alpha x-2\beta y.
 \end{gathered}
\end{equation}
This invertible transformation reduces
\eqref{eq:third-model} to
\begin{equation}\label{eq:third-intermediate-equation}
 w_{XXXY}+\frac12w_Yw_{XX}+w_Xw_{XY}
 +(1-\alpha)w_{XS}+\gamma w_{XZ}=0.
\end{equation}
The only remaining evolution direction is the constant
vector field
\begin{equation}\label{eq:third-vector-field}
 L=(1-\alpha)\partial_S+\gamma\partial_Z.
\end{equation}
If $L\neq0$, a linear change of coordinates rectifies it
to $\partial_T$, leaving a complementary coordinate $R$
absent from the differential operator.

\paragraph{Case $\gamma\neq0$.}
The combined transformation is
\begin{equation}\label{eq:third-combined-transformation}
 \begin{gathered}
 X=x,\qquad Y=y,\qquad T=\frac{z}{\gamma},\qquad
 R=t-y-\frac{1-\alpha}{\gamma}z,\\
 u(x,y,z,t)=v(X,Y,T,R)-\alpha x-2\beta y.
 \end{gathered}
\end{equation}
Its independent-variable Jacobian is $1/\gamma$, and it
maps \eqref{eq:third-model} to
\begin{equation}\label{eq:third-CBS-spectator}
 v_{XT}+v_{XXXY}+\frac12v_Yv_{XX}+v_Xv_{XY}=0.
\end{equation}
The inverse is
\begin{equation}\label{eq:third-combined-inverse}
 \begin{gathered}
 x=X,\quad y=Y,\quad z=\gamma T,\quad
 t=Y+R+(1-\alpha)T,\\
 v(X,Y,T,R)=u\bigl(X,Y,\gamma T,Y+R+(1-\alpha)T\bigr)
 +\alpha X+2\beta Y.
 \end{gathered}
\end{equation}
For each fixed $R$, equation \eqref{eq:third-CBS-spectator}
is precisely CBS. No condition $v_R=0$ has been imposed:
the absence of $R$-derivatives follows from an invertible
transformation of the full equation, not from a reduction
restricting its solutions.

\paragraph{Case $\gamma=0$, $\alpha\neq1$.}
The alternative transformation
\begin{equation}\label{eq:third-alternative-transformation}
 \begin{gathered}
 X=x,\qquad Y=y,\qquad T=\frac{t-y}{1-\alpha},\qquad R=z,\\
 u(x,y,z,t)=v(X,Y,T,R)-\alpha x-2\beta y
 \end{gathered}
\end{equation}
again gives \eqref{eq:third-CBS-spectator}. Its
independent-variable Jacobian is $-1/(1-\alpha)$; the inverse
is $x=X$, $y=Y$, $t=Y+(1-\alpha)T$, $z=R$, with
$v=u+\alpha X+2\beta Y$ at the corresponding arguments.
Hence, whenever
\begin{equation}\label{eq:third-generic-condition}
 (1-\alpha,\gamma)\neq(0,0),
\end{equation}
the third model is point-equivalent to CBS extended by a
spectator independent variable.

\paragraph{Case $\alpha=1$, $\gamma=0$.}
Here $L=0$, and
\eqref{eq:third-intermediate-transformation} gives
\begin{equation}\label{eq:third-degenerate}
 w_{XXXY}+\frac12w_Yw_{XX}+w_Xw_{XY}=0.
\end{equation}
Both $S$ and $Z$ are spectators. Thus, the exceptional
regime is the stationary CBS equation with two spectator
coordinates, whereas the generic regime is the CBS evolution
equation with one. Neither representation supplies a new
differential coupling in all four independent variables.

\section{Consequences for integrability, dispersion, and novelty}
\label{sec:consequences}

The equivalences established above explain several features
reported in \cite{Wazwaz2026ThreeCBS}. They also distinguish
properties inherited from classical CBS from claims that
would require additional justification.

\subsection{Inherited integrability properties}
\label{subsec:inherited-integrability}

An invertible point transformation allows local solutions
and auxiliary differential structures to be transferred
between equivalent equations. In particular, a Lax
representation for the canonical CBS equation can be
rewritten in the original coordinates, and conservation
laws and symmetry generators can be transported by the
corresponding transformation. For the first model and the
second model with $\alpha\neq0$, their relation to classical
CBS therefore supplies a direct explanation of the
integrability features observed in
\cite{Wazwaz2026ThreeCBS}.

For the third model under
$(1-\alpha,\gamma)\neq(0,0)$, this statement applies to
the CBS equation at each fixed value of the spectator
coordinate $R$. The resulting structures describe a
family of CBS systems parameterized by $R$; they do not,
by themselves, establish a new differential coupling
in that direction.

The transformations used here are nonsingular affine
maps in their respective parameter regimes. Consequently,
the local Laurent-series structure is transported without
introducing branching through the coordinate changes.
The CBS-type leading behavior and resonances reported in
the original article are therefore consistent with the
underlying equivalences. Their occurrence does not provide
independent evidence for a new integrable extension.

\subsection{Interpretation of the exponential dispersion relations}
\label{subsec:dispersion-interpretation}

The exponential dispersion relations can also be interpreted
through the point transformations.
For the first model, a phase
\begin{equation}
 \theta=kx+ly-\omega t+\delta
\end{equation}
becomes, under \eqref{eq:first-transformation},
\begin{equation}
 \theta=kX+lY-\Omega T+\delta,
 \qquad
 \Omega=\omega-\beta k-\alpha l.
\end{equation}
The canonical CBS exponential relation
$\Omega=k^2l$ therefore gives
\begin{equation}
 \omega=k^2l+\alpha l+\beta k,
\end{equation}
which is precisely the relation reported for the first
model. The additional terms are thus the frequency shift
associated with the translating coordinates.

For the third model, consider $\gamma\neq0$ and the phase
\begin{equation}
 \theta=kx+ly+mz-\omega t+\delta.
\end{equation}
Using the inverse transformation
\eqref{eq:third-combined-inverse}, this becomes
\begin{equation}\label{eq:third-transformed-phase}
 \theta=kX+(l-\omega)Y-\Omega T-\omega R+\delta,
 \qquad
 \Omega=(1-\alpha)\omega-\gamma m.
\end{equation}
The affine change of the dependent variable must also be
included: the original background $u=0$ corresponds to
\begin{equation}
 v_0=\alpha X+2\beta Y.
\end{equation}
Linearizing the canonical CBS equation about $v_0$,
with $v=v_0+\varepsilon q+O(\varepsilon^2)$, gives
\begin{equation}\label{eq:CBS-affine-linearization}
 q_{XT}+q_{XXXY}+\alpha q_{XY}+\beta q_{XX}=0.
\end{equation}
For an exponential perturbation with phase
\eqref{eq:third-transformed-phase}, equation
\eqref{eq:CBS-affine-linearization} yields
\begin{equation}
 \Omega=(k^2+\alpha)(l-\omega)+\beta k.
\end{equation}
Substituting the definition of $\Omega$ and rearranging
gives
\begin{equation}\label{eq:third-inherited-dispersion}
 (1+k^2)\omega
 =(k^2+\alpha)l+\beta k+\gamma m.
\end{equation}
This is exactly the exponential dispersion relation
reported for the third model.

Thus, the appearance of the parameters in this relation
is explained by the coordinate transformation and the
affine background. It does not establish a new dispersion
mechanism independent of classical CBS. Here, as in the
original exponential construction, the relation concerns
real exponential phases; it should be distinguished from
the oscillatory Fourier dispersion relation.

\subsection{Limits of the dimensional novelty}
\label{subsec:dimensional-novelty}

The third model retains four independent variables under
the invertible transformation. Accordingly, it is not
point-equivalent to a three-variable equation on a space
of lower dimension. The precise result is that it is
equivalent to the trivial extension
\begin{equation}
 \operatorname{CBS}[v(\,\cdot\,,R)]=0,
\end{equation}
in which $R$ is an additional independent variable absent
from the differential operator.

This distinction allows arbitrary parameter dependence
on $R$, subject to the CBS equation being satisfied for
each fixed $R$. It does not impose independence of $R$,
but it does show that the apparent fourth-direction
coupling is removable. Therefore, wave profiles drawn
in different coordinate planes, or phases containing
an additional wavenumber, do not alone demonstrate an
intrinsically four-variable integrable structure.

\subsection{Scope of the novelty assessment}
\label{subsec:scope-novelty}

The equivalence results do not invalidate exact solutions
merely because those solutions were obtained in noncanonical
coordinates. Nor do they establish, without a separate
comparison with the literature, that every reported solution
formula was previously known. They establish that the
proposed equations, in the generic regimes identified
above, do not define new point-inequivalent CBS structures.

Physical equivalence requires a separate assessment.
The transformations may interchange the original temporal
and spatial coordinates or introduce an affine background,
and therefore need not preserve specified initial data,
boundary conditions, or admissibility requirements.
A distinct physical interpretation could consequently
remain meaningful if supported by a concrete model and
a prescribed initial--boundary value problem. Such a
distinction, however, cannot be inferred solely from the
additional terms in the displayed equations.
\subsection{Recovery of the reported multi-wave solutions}
\label{subsec:recovery-multiwave}

The multi-wave constructions in \cite{Wazwaz2026ThreeCBS}
can be related explicitly to the earlier CBS solutions
in \cite[Sec.~3, pp.~594--595]{Wazwaz2008Multiple}.
The potential CBS equation considered in
\cite[Eq.~(17), p.~594]{Wazwaz2008Multiple} is
\begin{equation}\label{eq:earlier-CBS}
 w_{XT}+w_{XXXY}+4w_Xw_{XY}+2w_Yw_{XX}=0,
\end{equation}
where the transverse coordinate, denoted by $z$ in that
reference, has been renamed $Y$. The normalization $v=4w$
maps \eqref{eq:earlier-CBS} to
\eqref{eq:canonical-CBS}.

The logarithmic representation
$w=2(\ln f)_X$ in
\cite[Eq.~(24), p.~594]{Wazwaz2008Multiple}
therefore becomes
\begin{equation}\label{eq:earlier-normalized-solution}
 v=8(\ln f)_X.
\end{equation}
The one-, two-, and three-wave auxiliary functions,
given respectively in
\cite[Eqs.~(22), (26), and~(30), p.~594]{Wazwaz2008Multiple},
take the form
\begin{align}
 f_1&=1+e^{\Theta_1},\notag\\
 f_2&=1+e^{\Theta_1}+e^{\Theta_2}
       +A_{12}e^{\Theta_1+\Theta_2},\notag\\
 f_3&=1+\sum_{i=1}^{3}e^{\Theta_i}
       +\sum_{1\leq i<j\leq3}A_{ij}e^{\Theta_i+\Theta_j}
       +A_{12}A_{13}A_{23}
          e^{\Theta_1+\Theta_2+\Theta_3}.
 \label{eq:earlier-tau-functions}
\end{align}
Here,
\begin{equation}\label{eq:earlier-wave-data}
 \Theta_i=k_iX+L_iY-k_i^2L_iT+\delta_i,
 \qquad
 A_{ij}=\left(\frac{k_i-k_j}{k_i+k_j}\right)^2.
\end{equation}
The dispersion relation and pairwise interaction
coefficients are those of
\cite[Eqs.~(19) and~(28), p.~594]{Wazwaz2008Multiple};
the triple-interaction coefficient
$A_{12}A_{13}A_{23}$ is specified in
\cite[Eq.~(31), p.~594]{Wazwaz2008Multiple}.
The constants $\delta_i$ are included through translation
invariance. These formulas are considered where $f_N\neq0$,
with $k_i+k_j\neq0$ whenever $A_{ij}$ occurs.

\paragraph{Affine-background adjustment.}
For the second and third models, the affine terms in our
equivalence transformations must also be accounted for.
Direct substitution shows that, whenever $V$ satisfies
\eqref{eq:canonical-CBS},
\begin{equation}\label{eq:CBS-background-transformation}
 v(X,Y,T)
 =aX+2bY+V(X-bT,Y-aT,T)
\end{equation}
satisfies the same equation. Applying this identity to
\eqref{eq:earlier-normalized-solution} gives
\begin{equation}\label{eq:CBS-background-tau}
 v=aX+2bY+8\partial_X\ln f_N,
\end{equation}
where the auxiliary functions retain
\eqref{eq:earlier-tau-functions}, with phases
\begin{equation}\label{eq:CBS-background-phases}
 \Theta_i=k_iX+L_iY-\Omega_iT+\delta_i,
 \qquad
 \Omega_i=(k_i^2+a)L_i+bk_i.
\end{equation}
For the trivial extension by a spectator coordinate $R$,
the phase constants may be replaced by smooth functions
$\delta_i(R)$.

\paragraph{First model.}
Applying \eqref{eq:first-transformation} to
\eqref{eq:earlier-normalized-solution} gives
\begin{equation}
 u=8\partial_x\ln f_N,
 \qquad
 \Theta_i=k_ix+l_iy-\omega_it+\delta_i,
\end{equation}
where
\begin{equation}\label{eq:recovered-first-dispersion}
 L_i=l_i,
 \qquad
 \omega_i=k_i^2l_i+\alpha l_i+\beta k_i.
\end{equation}
Thus, the first model's multi-wave expressions follow
from the logarithmic representation and auxiliary
functions in
\cite[Eqs.~(24)--(31), pp.~594--595]{Wazwaz2008Multiple}
by the translating coordinate transformation.
The interaction coefficients $A_{ij}$ remain unchanged.

\paragraph{Second model.}
Assume $\alpha\neq0$. In
\eqref{eq:CBS-background-tau} and
\eqref{eq:CBS-background-phases}, choose
\[
 a=1,\qquad b=\beta,\qquad
 L_i=-\omega_i,\qquad \Omega_i=-\alpha l_i.
\]
Under \eqref{eq:second-transformation}, the background
$X+2\beta Y$ cancels the affine term $-x-2\beta t$,
yielding
\begin{equation}
 u=8\partial_x\ln f_N,
 \qquad
 \Theta_i=k_ix+l_iy-\omega_it+\delta_i.
\end{equation}
The background dispersion relation gives
\begin{equation}\label{eq:recovered-second-dispersion}
 \omega_i=\frac{\alpha l_i+\beta k_i}{1+k_i^2}.
\end{equation}
Consequently, the same auxiliary functions
\cite[Eqs.~(22), (26), and~(30), p.~594]{Wazwaz2008Multiple},
after the background adjustment and point transformation,
reproduce the second model's one-, two-, and three-wave
expressions, including their interaction coefficients.

\paragraph{Third model.}
First suppose $\gamma\neq0$, and use
\eqref{eq:third-combined-transformation}. Choose
\begin{equation}
 a=\alpha,\qquad b=\beta,\qquad
 L_i=l_i-\omega_i,\qquad
 \delta_i(R)=\delta_i^{0}-\omega_iR.
\end{equation}
The background-adjusted phases are then
\begin{equation}
\begin{aligned}
 \Theta_i&=k_iX+(l_i-\omega_i)Y
          -\Omega_iT-\omega_iR+\delta_i^{0},\\
 \Omega_i&=(k_i^2+\alpha)(l_i-\omega_i)+\beta k_i.
\end{aligned}
\end{equation}
In the original coordinates, these become
\begin{equation}
 \Theta_i=k_ix+l_iy+m_iz-\omega_it+\delta_i^{0},
\end{equation}
provided
\[
 \Omega_i=(1-\alpha)\omega_i-\gamma m_i.
\]
It follows that
\begin{equation}\label{eq:recovered-third-dispersion}
 \omega_i=
 \frac{(k_i^2+\alpha)l_i+\beta k_i+\gamma m_i}
      {1+k_i^2}.
\end{equation}
The background $\alpha X+2\beta Y$ cancels the affine
part of \eqref{eq:third-combined-transformation}.
Thus, the earlier logarithmic construction
\cite[Eq.~(24), p.~594]{Wazwaz2008Multiple} again yields
\begin{equation}
 u=8\partial_x\ln f_N,
\end{equation}
with the phases and interaction coefficients reported
in \cite{Wazwaz2026ThreeCBS}.

When $\gamma=0$ and $\alpha\neq1$, the same conclusion
follows from \eqref{eq:third-alternative-transformation},
using
\[
 L_i=l_i-\omega_i,\qquad
 \Omega_i=(1-\alpha)\omega_i,\qquad
 \delta_i(R)=m_iR+\delta_i^{0}.
\]
The exceptional regimes identified above are governed
instead by the stationary CBS equation and are not
covered by this recovery argument.

These identifications recover the one-, two-, and
three-wave constructions of
\cite[Eqs.~(24)--(31), pp.~594--595]{Wazwaz2008Multiple}
in the coordinates and normalization used in
\cite{Wazwaz2026ThreeCBS}, throughout the stated
nondegenerate regimes. The additional wavenumber in
the third model is accommodated by phase dependence
on the spectator coordinate, while the interaction
coefficients remain those of the earlier CBS solutions.
This comparison concerns the multi-wave families;
it does not establish prior publication of all the
other traveling-wave or cnoidal formulas in
\cite{Wazwaz2026ThreeCBS}.

\section{Conclusion}\label{sec:conclusion}

We have established explicit invertible point transformations
relating the three extended models proposed in
\cite{Wazwaz2026ThreeCBS} to the classical potential CBS
equation. The first model is equivalent to CBS for all
parameter values, and the second has the same property
when $\alpha\neq0$. For the third model, whenever
$(1-\alpha,\gamma)\neq(0,0)$, the transformed equation
is CBS with an additional spectator coordinate.
The exceptional regimes reduce to stationary CBS
equations with spectator coordinates.

Beyond equation equivalence, we have explicitly recovered
the reported one-, two-, and three-wave families from
the earlier CBS constructions in
\cite[Eqs.~(24)--(31), pp.~594--595]{Wazwaz2008Multiple}
and
\cite[Eqs.~(27)--(35), p.~175]{Wazwaz2010CBS}.
Normalization, the point transformations, and the
appropriate affine-background adjustment reproduce
their exponential phases and dispersion relations,
while preserving the interaction coefficients.
For the third model, the additional wavenumber is
accommodated by phase dependence on the spectator
coordinate. Thus, in the stated nondegenerate regimes,
these multi-wave families are transformed versions
of earlier CBS constructions.

The equivalences also explain the occurrence of CBS-type
integrability features. The additional terms in the first
model are removable constant drifts, while the apparent
fourth-direction coupling in the third model disappears
under an invertible transformation. Accordingly, neither
the proposed equations nor the recovered multi-wave
families establish new point-inequivalent CBS structures
in these regimes.

These conclusions do not invalidate the recovered
solutions or establish prior publication of every other
traveling-wave or cnoidal formula reported in
\cite{Wazwaz2026ThreeCBS}. Any remaining mathematical
novelty requires a separate demonstration beyond the
equation representations and multi-wave constructions
identified here. A distinct physical interpretation
likewise requires a specified initial--boundary value
problem, since the transformations need not preserve
the original temporal interpretation, boundary conditions,
or admissible backgrounds.

\section*{Statements and declarations}


\medskip
\noindent\textbf{Funding}
No external funding was received for this work.

\medskip
\noindent\textbf{Data availability}
All transformations supporting the findings are presented
in the manuscript.
No external datasets were generated or analysed.

\medskip
\noindent\textbf{Conflict of interest}
The authors declare no competing interests.

\medskip
\noindent\textbf{Ethics approval and consent to participate}
Not applicable.

\begin{thebibliography}{1}
\expandafter\ifx\csname url\endcsname\relax
  \def\url#1{\texttt{#1}}\fi
\expandafter\ifx\csname urlprefix\endcsname\relax\def\urlprefix{URL }\fi
\expandafter\ifx\csname href\endcsname\relax
  \def\href#1#2{#2} \def\path#1{#1}\fi

\bibitem{Wazwaz2026ThreeCBS}
A.-M. Wazwaz, W.~Alhejaili, S.~A. El-Tantawy, Three extended {Calogero--Bogoyavlenskii--Schiff} models in {$(2+1)$} and {$(3+1)$} dimensions: {Painlev{\'e}} integrability, multi-shock, cnoidal, and traveling waves, The European Physical Journal Plus 141 (2026) 1218.
\newblock \href {https://doi.org/10.1140/epjp/s13360-026-08459-x} {\path{doi:10.1140/epjp/s13360-026-08459-x}}.

\bibitem{Wazwaz2008Multiple}
A.-M. Wazwaz, Multiple-soliton solutions for the {Calogero--Bogoyavlenskii--Schiff}, {Jimbo--Miwa} and {YTSF} equations, Applied Mathematics and Computation 203~(2) (2008) 592--597.
\newblock \href {https://doi.org/10.1016/j.amc.2008.05.004} {\path{doi:10.1016/j.amc.2008.05.004}}.

\bibitem{Wazwaz2010CBS}
A.-M. Wazwaz, The (2+1) and (3+1)-dimensional {CBS} equations: Multiple soliton solutions and multiple singular soliton solutions, Zeitschrift f{\"u}r Naturforschung A 65~(3) (2010) 173--181.
\newblock \href {https://doi.org/10.1515/zna-2010-0304} {\path{doi:10.1515/zna-2010-0304}}.

\bibitem{LiChen2004}
B.~Li, Y.~Chen, Exact analytical solutions of the generalized {Calogero--Bogoyavlenskii--Schiff} equation using symbolic computation, Czechoslovak Journal of Physics 54~(5) (2004) 517--528.
\newblock \href {https://doi.org/10.1023/B:CJOP.0000024955.75594.8c} {\path{doi:10.1023/B:CJOP.0000024955.75594.8c}}.

\bibitem{olverbook}
P.~Olver, {Applications of Lie Groups to Differential Equations}, Vol. 107, Springer-Verlag Inc., New York, 1986.
\newblock \href {https://doi.org/10.1007/978-1-4612-4350-2} {\path{doi:10.1007/978-1-4612-4350-2}}.

\end{thebibliography}

\end{document}